\documentclass[a4paper]{spie}
\usepackage{graphicx}
\title{Sky coverage for Layer Oriented MCAO: a detailed analytical and 
numerical study}
\author{Carmelo Arcidiacono\supit{a}, Emiliano Diolaiti\supit{b}, Roberto 
Ragazzoni\supit{c,d}, Jacopo Farinato\supit{c}, Elise Vernet\supit{c}
\skiplinehalf
\supit{a}Dip.di Astronomia e Scienza dello Spazio-Univ. di Firenze,Largo 
E. Fermi 5, I-50125 Firenze,Italy; \\
\supit{b}INAF - Osservatorio Astronomico di Bologna, Via Ranzani 1, 
I-40127 Bologna, Italy;\\
\supit{c}INAF - Osservatorio Astrofisico di Arcetri, Largo E. 
Fermi 5, I-50125 Firenze, Italy;\\
\supit{d}Max-Planck-Institut f\"ur Astronomie, K\"onigstuhl 17, D-69117, 
Heidelberg, Germany
}
\authorinfo{Further author information: (Send correspondence to Carmelo Arcidiacono)\\E-mail: carmelo@arcetri.astro.it, Telephone: +39 049 829 3523\\              }

\begin{document}

 \maketitle

\begin{abstract}
One of the key-points for the future developments of the multiconjugate 
adaptive optics for astronomy is the availability of the correction 
for a large fraction of the sky. The sky coverage represents one of the 
limits of the existing single reference adaptive optics system. 
Multiconjugate adaptive optics allows to overcome the limitations due to 
the small corrected field of view and the Layer Oriented approach, in 
particular by its Multiple Field of View version, increases the number 
of possible references using also very faint stars to guide the adaptive 
systems. In this paper we study the sky coverage problem in the Layer 
Oriented case, using both numerical and analytical approaches. Taking 
into account a star catalogue and a star luminosity distribution 
function we run a lot of numerical simulation sequences using the Layer 
Oriented Simulation Tool (LOST). Moreover we perform for several cases a 
detailed optimization procedure and a relative full simulation in order 
to achieve better performance for the considered system in those 
particular conditions. In this way we can retrieve a distribution of 
numerically simulated cases that allows computing the sky coverage with 
respect to a performance parameter as the Strehl Ratio and to the 
scientific field size.
\end{abstract}

\keywords{Multi-Conjugate Adaptive Optics, Layer oriented 
MCAO, Wavefront Sensors, Sky Coverage}

\section{INTRODUCTION}

The theoretical and technological developments of multiconjugate 
adaptive optics\cite{beckers88,beckers89a,1994JOSAA..11..783E} (MCAO) give room to design a new generation 
instruments for 8-10 meters class telescopes. Moreover budget 
considerations induce to think more attractive ground based solutions 
than space ones, at least for the optical and near infrared portions of 
the electromagnetic spectrum. Actually large, very large and extremely 
large telescopes\cite{2003SPIE.4840..151D} projects take into account the MCAO option in order to achieve the theoretical diffraction limit resolution. However is still 
pending the way to approach to the MCAO correction. One of the biggest 
drivers for the selection of which MCAO technique should be used is the 
effective feasibility of the adaptive correction. For all natural guide 
stars techniques this becomes the probability to find (or not) a 
suitable stars asterism around the object, target of the scientific 
exposure. The probability to find these reference stars is what we call 
sky coverage. Laser guide stars techniques are quite solving this 
problem even if, to date, more technological efforts to achieve 
stability and high power for the laser-beam\cite{2003SPIE.4839..393R} are needed. Sky coverage 
usually is analysed in statistical way: assuming a star luminosity 
density function\cite{1980ApJS...44...73B} (for example the Bahcall and Soneira function) for the 
Galaxy the value of the coverage for different galactic latitudes and 
longitudes are then retrieved. 

In this paper we will discuss the sky coverage problem focusing on the 
Layer-Oriented\cite{LO2} (LO) MCAO technique and in particular on the Multiple 
Field of View\cite{mfov} (MFoV) version. This approach is very attractive because 
of the possibility to add optically the light of many reference stars 
sensed simultaneously on the same detector. In this way the signal to 
noise ratio (SNR) on the wave-front measurements does not depend on the 
brightness of the single reference star, but by the overall integrated 
magnitude of the asterism. This capability increases the sky coverage by 
including also faint stars in the list of the possible references. In LO 
each couple of deformable mirror (DM) and Wave-Front Sensor (WFS) is 
conjugated to a turbulent layer in an independent closed loop (from the 
hardware point of view). So, in principle, it is possible to drive a 
MCAO system using different reference stars for each detector-DM loop.

The MFoV approach goes in the direction of increasing the sky coverage 
using for the ground layer correction a large field of view (FoV) of 6 
arcmin or more, instead of the typical 1 or 2 arcmin (with a factor 16 
gain in terms of sky area) where to look for the reference stars. The 
ground layer usually is the strongest contributor to the atmospheric 
seeing and if it is subtracted then the residual phase to be corrected 
by the other (or others) high loop is considerably smaller. High layers 
have usually Fried parameter (r$_{0}$) larger than the ground 
one and a bigger coherence time ($\tau_{0}$). One the advantages of the 
LO is the possibility to tune the characteristics of the loops as the 
integration time and the dimension of the sub-apertures, that compose 
the measured wave-front, to the statistical properties of the turbulent 
layers. In this case the MFoV systems, or more generally the LO ones, 
take advantage of this characteristic driving the high layer with a 
frequency usually different than the ground (because the turbulence is 
usually lower in the high layers than in the ground one and in the MFoV 
case also the reference stars are different between the two conjugation 
altitudes) optimising in this way the signal to noise ratio for two 
WFSs. In the system we are going to analyse in this study the reference 
stars for the ground layer correction are chosen in a ring of 6arcmin 
external diameter excluding the inner corrected FoV of two arcmin, that 
is the field used for the selection of the guide stars for the high WFS.

In 2002 we started the development of a numerical simulation code, then 
called LOST - Layer Oriented Simulation {Tool\cite{2004ApOpt..43.4288A}}, to analyse the performance 
of the MCAO system using the LO approach. It is very close to being an 
end-to-end simulation code: the main difference is in the way the WFSs 
are simulated. In LOST the phase noise introduced by each detector is 
computed through the relation derived for the Shack-Hartmann WFS and then 
it is added to the noise-free measurements. 

In this paper we will describe the USNO {catalogue\cite{2003AJ....125..984M}}, the definition of sky 
coverage and the adaptive system we considered in this study. Finally 
the data are discussed.

\section{ Sky coverage estimation using stars catalogue}
\subsection{The USNO B catalogue}
Every sky-coverage study is strictly dependent on the initial parameter 
taken into account for the adaptive system being considered, and 
dramatically on the way to select the reference stars. We analysed three 
1 square degree real sky-fields, reading the stars data in the USNO-B 
catalogue (version 1.0). In this catalogue are listed the objects 
identified by scanning plates taken in the last 50 years by different 
observatories. These data cover all the galactic latitudes and 
longitudes and the catalogue gives information about positions, proper 
motions, magnitudes in various optical pass bands (B, R and I), and a 
star/galaxy estimator for more than 1 billion objects. This catalogue 
has an accuracy of about 0.2 arcsec for astrometry, 0.3 arcsec in 
photometry and 85 percent in star/galaxy classification. It is supposed 
to be completed until the magnitude 21 in the V band.

\begin{figure}
\centerline{\includegraphics[width=2.8in]{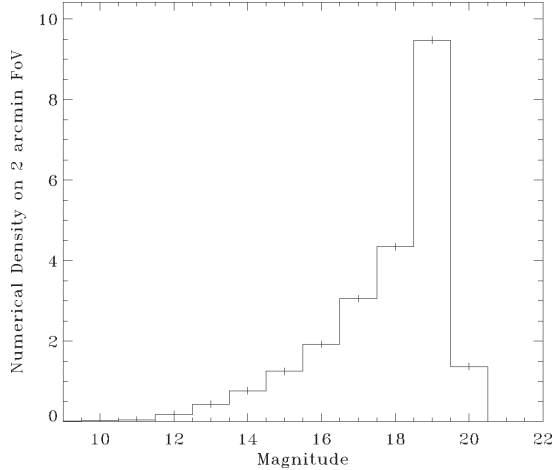}}
\caption{\footnotesize{This histogram shows the distribution of the numerical density of the stars in the catalogue with respect to the magnitude in the direction of the Galactic anti-center. The numerical density represents the average number of stars over 2 arcmin FoV for each unit magnitude range. It is clear that the catalogue is not complete for stars fainter than $19^{\rm th}$ R-magnitude.}}\label{fig:1}
\end{figure}

 However the study we are presenting refers to the R band only, and we 
checked that the catalogue presents lack of data for stars fainter than 
19$^{th}$ magnitude in this band (see Figure~\ref{fig:1}).

\subsection{Sky Coverage}\label{section:_Ref76797586}

The Sky Coverage is defined as the fraction of sky where the MCAO 
correction is feasible. However this concept suffers of a lot of 
ambiguities. These can be generated by the definition itself. In fact to 
assess that MCAO correction it is possible one can refer to simple 
closure of the adaptive loop with, for example, a small gain in terms of 
SR or Full Width Half Maximum of the PSF, or he can refer to the level 
of SR that an astronomer define as good for his purposes, or taking into 
account encircled energy or other details. Of course this definition 
depends on the target of MCAO correction and on the system 
characteristics (for example the performance requested for a ground 
layer corrector are different than a Multi-conjugate system). 

Someone\cite{LO2,2003SPIE.4839..566M} defines sky-coverage as the percentage of cases where 
the SR achieve at least the 50\% of the infinite Signal to Noise Ratio 
(SNR) case. But this definition is not regarding at all the absolute 
performance of the system, and in several cases the obtained results 
could not be enough high to give the requested correction (because of 
very low SR or big un-uniformity in the correction) but included in the 
list of the sky-coverage as good.

We propose to relate the sky-coverage definition to the astrophysical 
target to be study without speaking of ``absolute sky coverage". This 
means that we have to set in some way a condition parameter to 
distinguish the region of the sky where there is (or not) the coverage. 
Looking to the system we are considering (MCAO on a 8 meter telescope) a 
reasonable parameter can be the SR, and the lower limit could be 5 or, 
better 10 percent (at the K band), useful for many astrophysical 
studies. Here we considered different classes of objects according to 
their angular dimensions so we defined different coverage for 3 
different FoVs, averaging the SR data over the interested region. In 
this way we take into account also un-uniformities of the correction 
that were not considered if we only take as representative the maximum 
SR over the field taken into account.

The sky-coverage problem regarding the LO-MFoV technique has been 
studied\cite{2003SPIE.4839..566M} using a statistical approach based on the Bahcall and Soneira 
star density function and an analytical model for the Pyramid {WFS\cite{pyramid,pyrmodal}}. Here the limiting integrated magnitude to have a 
degradation of 0.5 in SR with respect to the case with infinite bright 
reference stars are 14.5 for the ground loop and 16.5 for the high loop. 
There were found out sky coverage between 8\% and 20\% for the galactic 
poles and between 46\% and 96\% for low galactic latitudes. 

Here we consider the definition of coverage we proposed few lines above 
and this method: first of all we consider real 1 square degree fields, 
from the USNO-B catalogue, around the Galactic poles, North (NGP) and 
South (SGP) and the Galactic anti-Centre (GAC) and then we define on 
them a square grid of $32\times32$ points with a step of 101 arcsec. Every grid 
point represents the on-axis direction where the FoV circles of 2 and 6 
arcmin are centred. 

But in the USNO-B catalogue are listed also extended non-stellar objects 
that cannot be references for MCAO: we discard these objects from the 
data by considering the star/galaxy estimator of the catalogue. In this 
way we have the ``raw" data for the analysis of the field stars 
distribution and characteristics. 

Of course the sky-coverage is limited by the presence of the stars in 
the field of view but also by other very important parameters. The 
reference stars are selected (if any) simultaneously in the two FoVs of 
2 and 6 arcmin and according to the technical constrains summarized in 
Table~\ref{table:1}. The constrain on the maximum range of 
variation for the references brightness usually tends to decrease the 
probability to find very bright star (brighter than 12-13 mag) in the 
central 2 arcmin FoV because the other stars in the Field are 
``statistically" fainter than 3 magnitudes. Another limit to the 
sky-coverage is the minimum distance that separates two close 
references. This is due to the physical dimension of the mounting of the 
pyramid (also called star enlargers). These values depend on the system, 
in particular we used numbers taken from an existing {project\cite{2003SPIE.4839..536R}}: the 
minimum separations adopted are 20 arcsec and 30 arcsec respectively for 
the high and for the ground WFSs.

The stars are selected in order to retrieve the brightest as possible 
asterism considering the limits given and trying to maximize the 
separation between the references.

\begin{table}[h]
\begin{center}
\begin{tabular}
{|p{90pt}|p{50pt}|p{50pt}|p{60pt}|p{60pt}|p{45pt}|p{45pt}|}
\hline
\rule[-1ex]{0pt}{3.5ex}  Limit integrated magnitude for the asterism & Min Separation GL & Min 
Separation HL & Max magnitude range & Min NGS number GL and HL & Max NGS 
number GL & Max NGS number HL \\
\hline
\rule[-1ex]{0pt}{3.5ex}  R $<$ 19 & 30 arcsec & 20 arcsec & 3 mag & 3 & 12 & 8 \\
\hline
\end{tabular}
\caption{\footnotesize{In this table are listed the main 
constrains used for the selection of the Natural Guide Stars. All this 
condition must be verified simultaneously in order to select the stars 
asterism for a simulation run. If no asterism verifies these conditions 
then in that region we say that there is not sky coverage.}}\label{table:1}
\end{center}
\end{table}

A simulation run is performed for each grid-point where a suitable 
asterism has been found, retrieving in this way a grid of SR values on 
the 2 arcmin central FoV (see Figure~\ref{fig:2}). Finally the simulations outputs are the 
data used for the sky-coverage analysis.

\subsection{The system taken into account}
In order to compare the results obtained on different asterisms of the 
same 1 degree field the characteristics of the LO system and of the 
atmosphere must be fixed. We consider an 8-meter telescope and the 
median turbulence profile typical of the Paranal (see Table~\ref{table:4}), characterized by an average seeing value at 
the V band of 0.73 arcsec equivalent to an overall r$_{0}$ of 0.14 
meters (in the V band). In both cases the results are computed in the K 
band, $2.2\mu m$, with a correspondent r$_{0,K}$=0.83 meters. We consider 
a MFoV system with 2 DMs conjugated to 0 and 8.5~km. The spatial 
geometry of the system has been taken fixed to a sampling $8\times8$ for the 
ground and $7\times7$ for the high in order to be deep in term of limiting 
magnitude but not optimising for the maximum achievable SR. The other 
mains MCAO system parameters are listed in Table~\ref{table:2}. Each 
simulation covers an elapsed time of 0.5 seconds, but only the last 0.4 
are taken into account for the computation of the long exposure SR.

\begin{table}[h]
\begin{center}
\begin{tabular}{|p{41pt}|p{45pt}|p{45pt}|p{40pt}|p{50pt}|p{35pt}|p{50pt}|p{35pt}|p{40pt}|}
\hline
\rule[-1ex]{0pt}{3.5ex} Overall efficiency & Sensing wavelength & Scientific wavelength & 
Bandwidth & Conjugation altitude & RON \textit{e}/frame & Dark 
Current & Binning & Max \# Zernike modes \\
\hline
\rule[-1ex]{0pt}{3.5ex} 0.2 & 0.7$\mu m$ & 2.2$\mu m$ & 0.4$\mu m$ & 0 km & 3.0 & 200 e$^{-}$/sec & 2$\times$2 & 59 
\\
\hline
\rule[-1ex]{0pt}{3.5ex}  & & & & 8.5 km & 3.0 & 200 e$^{-}$/sec & 4$\times$4 & 45 \\
\hline
\end{tabular}
\caption{\footnotesize In this table are presented the main characteristic of the MCAO system considered. We simulate an MFoV system with corrected field of 2 arcmin and a technical FoV of 6 arcmin relative to the high and ground WFSs respectively.}\label{table:2}
\end{center}
\end{table}

The integration times applied to the two WFSs are tuned to the 
integrated magnitude on the 6arcmin ring and on the 2arcmin FoVs for the 
ground and the high respectively. As in the case of the spatial 
sampling, we choose the frame rates of the two WFSs to retrieve a high 
limiting magnitude instead of a higher SR. The values of the integrated 
times are listed in Table~\ref{table:3} below.

\begin{table}[h]
\begin{center}
\begin{tabular}{|l|l|l|l|l|l|}
\hline
Ground Loop & R$_{int}$ $<$ 11 & 11 $<$ R$_{int}$$<$13 & 13 $<$ R
$_{int}$$<$15 & 15 $<$ R$_{int}$$<$16.5 & R $>$ 16.5 \\
\hline
 & 2 msec & 4 msec & 10 msec & 20 msec & 40 msec \\
\hline
High Loop & R$_{int}$ $<$ 10 & 10 $<$ R$_{int}$$<$12 & 12 $<$ R
$_{int}$$<$14 & 14 $<$ R$_{int}$$<$16.5 & R $>$ 16.5 \\
\hline
 & 2 msec & 4 msec & 10 msec & 20 msec & 40 msec \\
\hline
\end{tabular}
\caption{\footnotesize In this table are listed the 
integration times used for the two WFS with respect to the integrated 
magnitude of the both references asterism. The values are tuned to the 
statistical characteristics of the conjugated planes. }\label{table:3}
\end{center}
\end{table}

\begin{table}[h]
\begin{center}
\begin{tabular}{|l|l|l|l|}
\hline
Layer ID & Layers Altitude $[$m$]$ & Cn2 fraction & Wind $[$$^{m}$/
$_{s}$$]$ \\
\hline
1 & 0 & 0.65 & 6.6 \\
\hline
2 & 1800 & 0.08 & 12.4 \\
\hline
3 & 3200 & 0.12 & 8.0 \\
\hline
4 & 5800 & 0.03 & 33.7 \\
\hline
5 & 7400 & 0.03 & 23.2 \\
\hline
6 & 13100 & 0.08 & 22.2 \\
\hline
7 & 15800 & 0.01 & 8.0 \\
\hline
\end{tabular}
\caption{\footnotesize Here are listed the atmospheric 
parameters used in the simulations. For each layer an outer-scale of 20 
m has been considered. The isoplanatic angle for the overall atmosphere 
is about 15 arcsec at the 2.2$\mu m$ pass band. This model is not the most 
recent one where there is a bit more turbulence power in the ground 
layer (67\% instead of 65\%). In this study a 0".73 seeing in V Band and 
0".66 seeing in R band were considered.}\label{table:4}
\end{center}
\end{table}

\subsection{Optimization test}
The integration times used for both loops were set according to the 
integrated magnitude of the asterism in the 6 arcmin annular FoV and in 
the central 2 arcmin FoV, respectively for the Ground and the High WFS 
(see Table~\ref{table:3}). This solution to set this important 
couple of parameters is correct only for a first order approach. In 
fact, for example, it neglects the effect of the different illumination 
of the sub-apertures in the High WFS due to the references position and 
different brightness. A smarter analysis should take into account a 
fine-tuning of the frame rates for the two WFSs. But an optimization 
procedure to be applied to all the asterisms considered it's not 
feasible because in this case the overall number of simulations 
performed will increase too much. In other cases described so far\cite{LNPDRLOSDA,MADLOSDA} we considered a grid of possible values for the two frame rates in 
order to taking into account the different combinations of the two 
integration times. Here we considered a small $3\times3$ grid with values 
around to the ones specified in Table~\ref{table:3} (these 
depending on the integrated magnitude) ranging from 25$\%$ less and 25$\%$ 
more the two values taking into account for the simulation performed 
yet. We optimized the integration time only for a small set of the 
asterisms (20) used in the three 1$\times$1 square degree fields. The results 
of this optimization compared to the non-optimized data allow 
extrapolating the optimized SR values for all the simulated cases. 

\section{Data analysis}
\subsection{CPU time and Workstation}
We found in the catalogue 40000 useful stars over the 3 sky-fields 
considered. We analysed 3072 regions, running 2000 simulations on the 
found asterisms. Each simulation took 3-6 hours of CPU time according to 
the CPU clock (at most we used 2000 MHz). 

The overall CPU time used was of 330days, the most spent on the Arcetri 
Beowulf cluster with 16 nodes, each one equipped with two CPUs. The LOST 
code is not parellelized yet, but we ran ``in parallel" different 
simulations on different nodes at the same time.

\begin{figure}
\centerline{\includegraphics[width=2.8in]{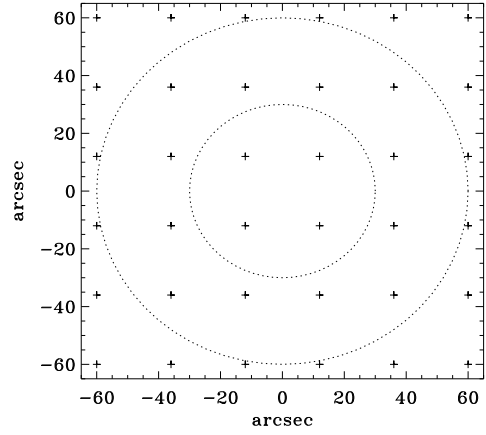}}
\caption{\footnotesize{In this plot the ``+" sign defines the direction where the SR is computed in all the simulated cases, the step of this square grid is 24arcsec. The two circles represent the 1arcmin field and the 2arcmin. The correction was applied in the 2 arcmin FoV. In addition to these $6\times6$ directions the SR data were computed also in the Natural Guide Stars positions.}}\label{fig:2}
\end{figure}

\begin{figure}
\centerline{\includegraphics[width=4.0in]{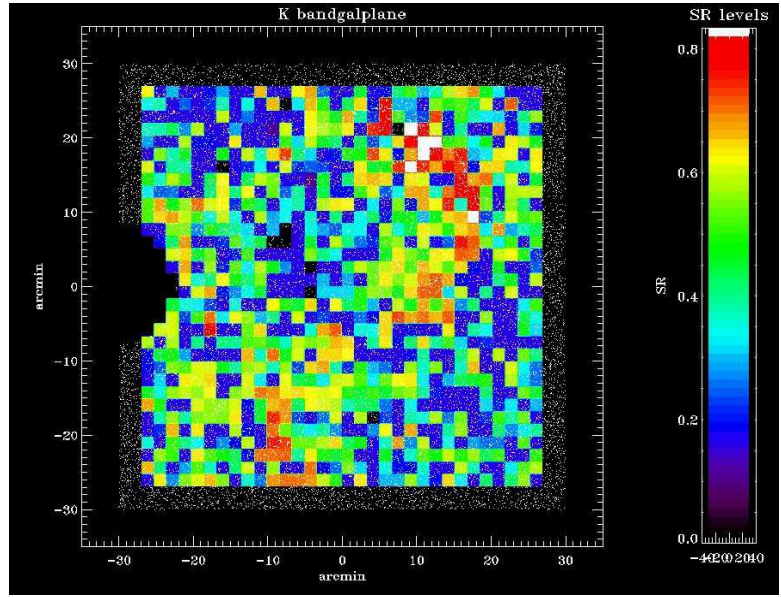}}
\caption{\footnotesize{This picture shows a map of the analytical SR found for the Galactic Plane case. Each square covers 101$\times$101arcsec region. If there is a lack of data or no good asterisms to drive the adaptive system this square is black. The color bar on the right indicates the value of the SR. For example: on the left of this map there is a very bright star that saturates all the plates taken into account for the catalogue preparation: in fact a circular hole without stars appears in our analysis. The analytical SR gives only an idea of the possible SR achievable because it does not take into account neither the distribution of the stars in the 6 and 2 arcmin FoV nor the turbulence profile.}}\label{fig:3}
\end{figure}

We developed an IDL procedure to manage the different steps described in 
the section above. Moreover this found the best asterism for each couple 
of 2arcmin and 6arcmin fields; the best integration times for the two 
loops and it computed an analytical SR over the 3 1$\times$1 deg$^{2}$ 
fields considering the asterism and integration times found before.

\subsection{Data analysis}
For each simulation LOST computed the SR values over the 2arcmin FoV. A 
6$\times$6 grid of SR evolution data was retrieved for each of the 32$\times$32 positions in the three 1-degree field (see Figure~\ref{fig:2}). Using these data the averages SR were computed on the 1arcmin circle 
and the 2-arcmin FoV. Moreover the on axis direction SR was taken into 
account averaging the 4 ``probe" stars close to the centre of the field 
(see Figure~\ref{fig:2}).

We optimize the MCAO system parameters looking for the loops closure and 
the robustness of the correction and we do not optimize the system in 
order to achieve high SR (more than 60\%). In fact considering only the 
effect of the spatial sampling of the metapupils for the ground (8$\times$8) 
and for the high loop ($7\times7$) the best SR achievable is between 55\%-60\% 
with the atmospheric parameters here considered.

Each simulation performed has an iteration step of 2 msec for a total of 
250 iterations that gives an overall time of 0.5 seconds. The length of 
these simulations is not enough to estimate a representative long 
exposure SR because of the effect of the bootstrap, so we assumed the 
maximum SR achieved during the run as reference for our analysis 
(the SR values we consider take into account the tip-tilt residual 
also). Analyzing the SR data of the 3 different 1 square degree fields 
considered, we drew different SR maps according to the 3 FoV sizes we 
assumed. These cases can be seen as representative of different 
instruments:
\begin{itemize}
\item A camera with few arcsec FoV;\\
\item An instrument with 1 arcmin FoV;\\
\item An instrument, or more instruments mounted on the same system 
covering the corrected 2arcmin field.
\end{itemize}

\subsubsection{Few arcsec FoV Case}\label{section:a3b2c1}
We assumed mounted at the focus of the telescope a camera with a small FoV of few arcsec centered in the optical axis direction. In this case the SR must be uniform because the FoV is smaller than the isoplanatic patch size. So we considered representative for this case the value of the SR obtained for the 4 ``probe" stars more close to the centre of the FoV (see Figure~\ref{fig:2}). Figure~\ref{fig:4} presents the results in terms of sky coverage VS threshold SR:
\begin{figure}
\centerline{\includegraphics[width=3.6in]{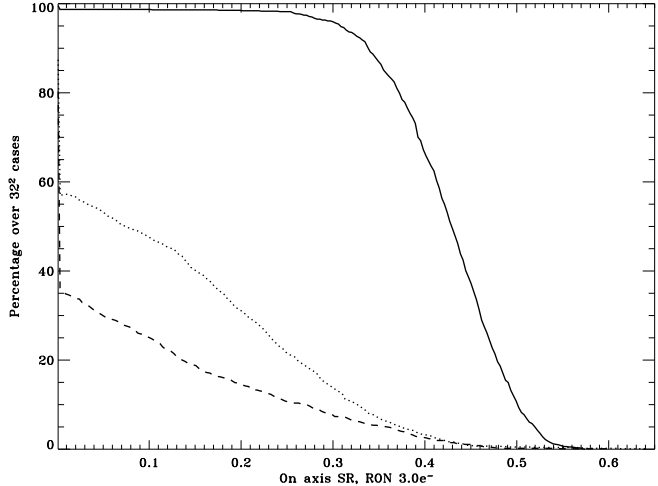}}
\caption{\footnotesize{This picture shows the sky coverage results for the three galactic latitude cases taken into account and relative to the on axis direction only. The functions plotted here represent the percentage of the simulated case where at least the SR showed in abscissa was achieved. Dotted line represents the North Galactic Pole; the dashed one refers to the South Galactic Pole and the solid to the Galactic Anti-centre. The percentage is relative to the 32$\times$32 directions considered for each galactic field.}}\label{fig:4}
\end{figure}
\begin{figure}
\centerline{\includegraphics[width=3.6in]{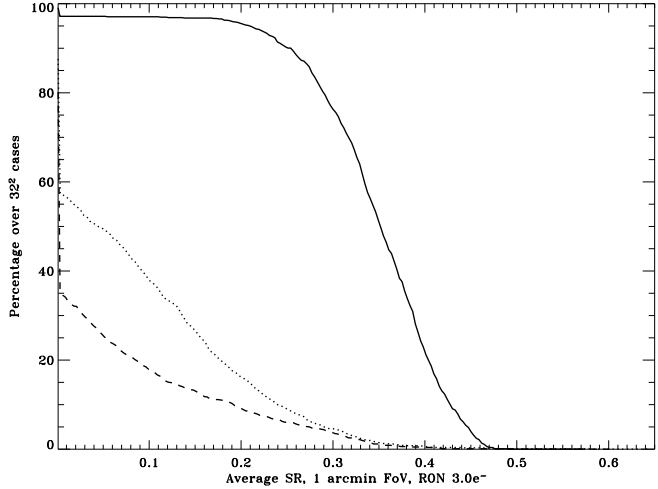}}
\caption{\footnotesize{This picture shows the results for the central 1arcmin FoV case for the three galactic latitudes taken into account. The functions plotted here represent the percentage where at least the SR showed in abscissa was achieved. Dotted line represents the North Galactic Pole; the dashed one refers to the South Galactic Pole and the solid to the Galactic Anti-centre.}}\label{fig:5}
\end{figure}

In the 98$\%$ of the cases taken into account the SR on axis was higher 
than 10$\%$, while percentages of 48$\%$ and 25$\%$ were retrieved 
respectively for the North and the South Galactic poles. For the low 
Galactic latitudes in half of the cases considered the SR on axis was 
higher than 40\%. 

\subsubsection{1 arcmin FoV case}\label{section:a3b2c2}
Now we describe the sky coverage analysis for an instrument with 1arcmin 
FoV centered in the axis direction (the same axis relative to the 
correction applied by the adaptive system). For this case the 
representative SR is the average SR over the central one arcmin computed 
by the simulations. The Figure~\ref{fig:5} shows the results 
relative to this field size: for the low galactic latitude (the Galactic 
Anticentre) in the 98\% of the directions considered the average SR was 
higher than 10\%, while the same values for the North and South Poles 
were 38\% and 17\% respectively.

\subsubsection{2 arcmin FoV case}\label{section:a3b2c3}
In this last case we supposed several instruments (or a unique big 
camera) observing the whole region corrected by the MCAO system (we 
considered a corrected FoV of 2 arcmin). As in the 1arcmin case we took 
into account the average SR, but now over the 2arcmin FoV. 

In the the results are presented: considering a 10\% threshold for the 
SR as condition to define the coverage we found sky coverage of 99\% for 
Galactic plane, 25\% and 13\% respectively for the North and South Galactic poles. 

\begin{figure}
\centerline{\includegraphics[width=3.6in]{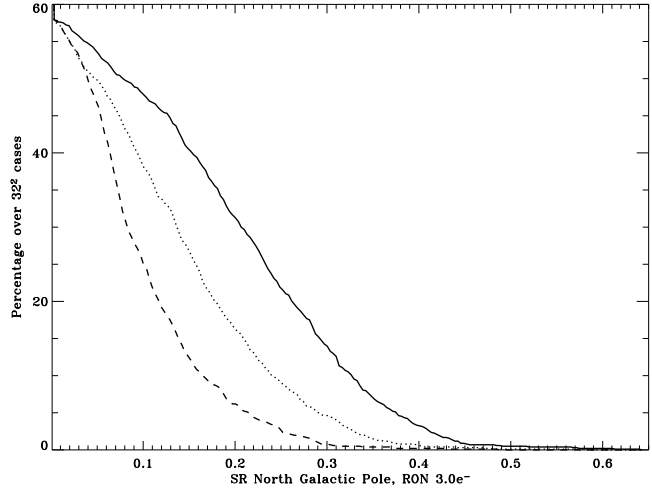}}
\caption{\footnotesize{This figure shows the percentage of sky-coverage with respect to a threshold SR. Dashed line represents the coverage with respect to the on axis direction SR; the dotted one refers to the average SR over 1 arcmin FoV and the solid line to the SR averaged over the 2arcmin corrected field. All the 3 curves refer to the analysis performed on the 1$\times$1 square degree field centered in the North Galactic Pole.}}\label{fig:7}
\end{figure}

We want to stress that the sky-coverage values within the 3 FoV sizes 
changes of a factor $\sim 2$ for the galactic poles while it is un-changed 
for the Galactic anticentre (see Figure~\ref{fig:7}). This 
different behaviour depends on the different stars density of the two 
galactic regions. The poles are poor of stars with respect to the low 
galactic latitudes: this translates in a less number of reference stars 
for the galactic poles and so a lower uniformity for the correction with 
respect to the galactic plane where it is quite easy to cover 
homogenously the corrected 2arcmin FoV with natural guide stars. 

\begin{figure}
\centerline{\includegraphics[width=3.6in]{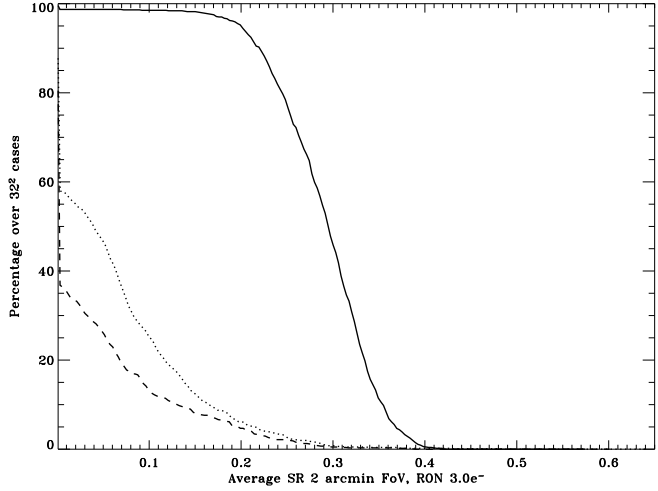}}
\caption{\footnotesize{This picture shows the results in the overall corrected 2 arcmin FoV and for the three galactic latitude cases analyzed. The functions plotted here are representing the percentages where at least the SR showed in abscissa was achieved. Dotted line represents the North Galactic Pole; the dashed one refers to the South Galactic Pole and the solid to the Galactic Anti-centre. The SR here considered is average SR over the corrected field of 2 arcmin.}}\label{fig:6}
\end{figure}

\subsubsection{Dealing to a different definition}
Now we want to analyse the results presented in the previous sections 
(\ref{section:a3b2c1}, \ref{section:a3b2c2} and 
\ref{section:a3b2c3}) according to a different definition of sky 
coverage, using, for example, that one given in the {reference\cite{2003SPIE.4839..566M}}, and that 
we discussed also above (section~\ref{section:_Ref76797586}). We assume 
as reference for the infinite SNR SR the maximum SR achieved in each of 
the 3 cases of Field of View considered that are: $\sim$0.6 for the few 
arcsec field of view case, $\sim$0.5 for the 1 arcmin case and $\sim$0.4 for the 
2 arcmin FoV. Using these values we found the SR$_{50}$ thresholds 
(50\% of the SR relative to the infinite SNR case) for each of these 
cases: 0.3, 0.25 and 0.2 respectively. Applying these thresholds instead 
of the 10\% one, we found coverage of 90\% for the low galactic 
latitudes case and between 7\%and 15\% for the galactic poles. Even if 
referring to different wavelengths these values agree with the ones 
given in the references\cite{mfov,2003SPIE.4839..566M} (here we considered correction in the K 
band while it was the R band in the {reference\cite{2003SPIE.4839..566M}}).
\begin{figure}
\centerline{\includegraphics[width=4.0in,height=3.2in]{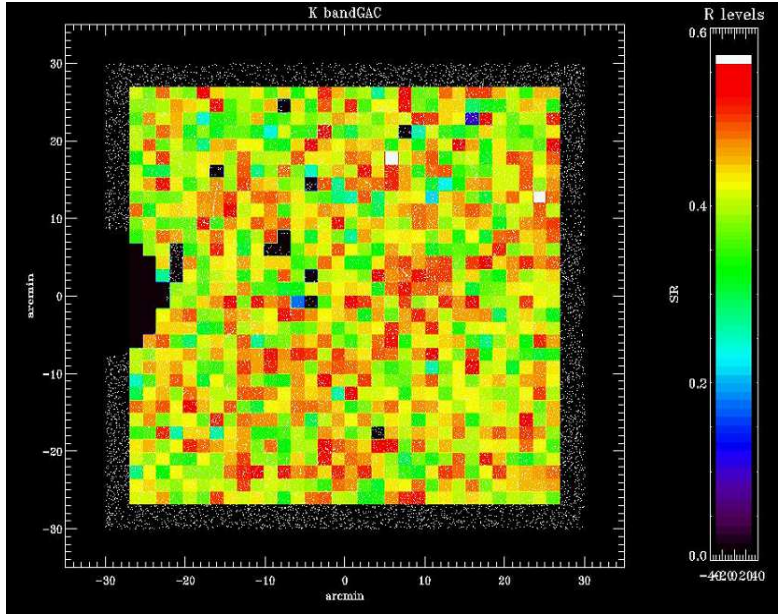}}
\caption{\footnotesize{This figure shows the results for the Galactic Anti-Centre in the on axis case. Different colors indicate different SR values.}}\label{fig:8}
\end{figure}

\begin{table}[h]
\begin{center}
\begin{tabular}{|l|l|l|l|}
\hline
 & North Galactic Pole & South Galactic Pole & Galactic Anticentre \\
\hline
On Axis(Few arcsec FoV) & 48 \% & 25 \% & 99 \% \\
\hline
1 arcminFoV & 38 \% & 17 \% & 97 \% \\
\hline
2 arcminFoV & 25 \% & 13 \% & 99 \% \\
\hline
\end{tabular}
\caption{In this table are summarized the results 
for the different cases analyzed. A limit SR of 10\% was assumed.}\label{table:5}
\end{center}
\end{table}

\section{Example of SCIENCE-COVERAGE: cluster of galaxies at high 
red-shift }

What we said and stressed about different sky-coverage for different FoV 
becomes important when we consider the possible astronomical 
applications. For example we took as possible astronomical target the 
cluster of galaxies. As everybody knows their apparent dimension changes 
with their distance, but because of the structure and evolution of the 
universe this depend also by cosmological parameters. 

\begin{figure}
\centerline{\includegraphics[width=4.0in]{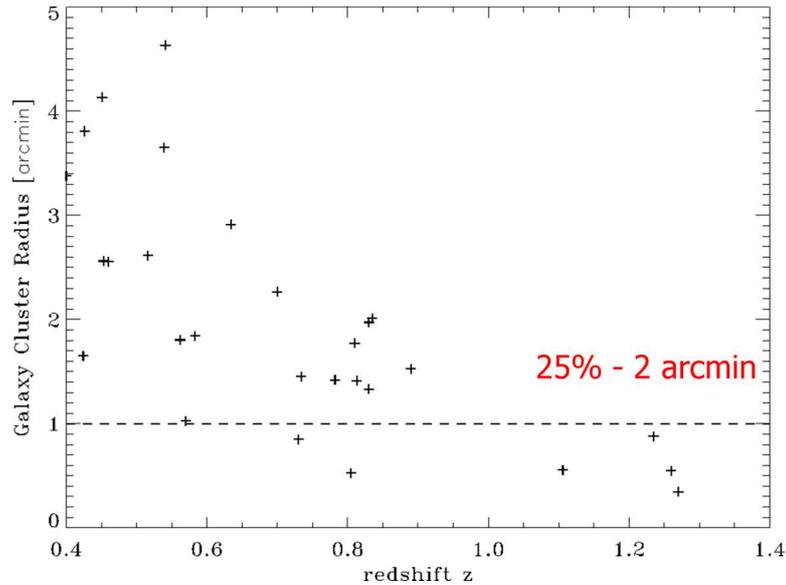}}
\caption{\footnotesize{This figure shows the angular size with respect of the red-shift according to the data in the {reference\cite{2004A&A...417...13E}}. We plotted a dashed line to a radius of 1 arcmin, corresponding to the 2arcmin FoV case we considered in the sky-coverage analysis. Following our results the clusters at $z \sim0.9$ have sky coverage of 25\% (at North Galactic Pole).}}\label{fig:9}
\end{figure}

The angular size of clusters is related to the red-shift (z) then for 
these objects the sky-coverage is a function of the z. We plotted the 
angular size taken from the reference with respect to the redshift\cite{2004A&A...417...13E} in 
Figure~\ref{fig:9}.

This plot says that high-z clusters have higher sky-coverage. If the 
cluster has dimension bigger than 2~arcmin then more 
contiguous asterisms are needed  to cover its entire dimension and, in this case sky 
coverage decreases. 

\section{Conclusions}

We analyse the sky coverage problem in the case of a specific Layer 
Oriented Multiple Field of View system. We showed basic relationship 
between sky coverage and field dimension to be studied and presented a 
scientific case. We analyse the definition of sky coverage and we 
stressed that it must be related to the performance requested to the 
adaptive system and the class of objects to be studied with the 
scientific instrument to be used. In particular we set a reasonable 
threshold to the 10\% in SR as condition to define if there is sky 
coverage or not. We showed that for low galactic latitudes the 
correction is feasible about everywhere while at the galactic poles the 
coverage decreases, but down to reasonable values (20\%-40\%) to justify 
the use of this natural guide stars technique also for high galactic 
latitudes targets.

\section*{ACKNOWLEDGEMENTS}

Thanks to M. Le Louarn for the atmosphere profile of the Cerro Paranal 
(Chile), and to A. Puglisi as ``problem-solver" regarding the Beowulf 
cluster.

\bibliography{spie}   
\bibliographystyle{spiebib}   

\end{document}